\documentclass[twocolumn,showpacs,preprintnumbers,amsmath,amssymb,superscriptaddress]{revtex4}
\usepackage{amsfonts,amssymb,bm,graphicx,times} \usepackage{amsmath}
\usepackage{float,color}

\begin{document}

\title{Random unitary dynamics of quantum networks}

\author{J. Novotn\'y}
\affiliation{Institut f\"ur Angewandte Physik, Technische
Universit\"at Darmstadt, D-64289 Darmstadt, Germany}
\author{G. Alber}
\affiliation{Institut f\"ur Angewandte Physik, Technische
Universit\"at Darmstadt, D-64289 Darmstadt, Germany}
\author{I. Jex}
\affiliation{Department of Physics, Czech Technical University in Prague, 
115 19 Praha 1 - Star\'e M\v{e}sto, Czech Republic}

\begin{abstract}
We investigate the asymptotic dynamics of 
quantum networks under repeated applications of random unitary operations.
It is shown that in the asymptotic limit of large numbers of iterations this dynamics 
is generally governed by a typically low dimensional 
attractor space. This space is determined completely
by the unitary operations involved and it is independent
of the probabilities with which these unitary operations are applied.
Based on this general feature analytical results are presented for the asymptotic dynamics of 
arbitrarily large cyclic qubit networks whose nodes are
coupled by randomly applied controlled-NOT operations. 
\end{abstract}

\maketitle
%%%%%%%%%%%%%%%%%%%%%%%%%% classical networks into the quantum domain %%%%%%%%%%%%%%%%%%%%%%%%%%%%%%%%%%%%%%%%%%%%
In recent years specific features of classical networks have been investigated intensively \cite{Barabasi} because
they exhibit generic organizing principles shared by rather different systems, such as living cells or the internet.
In view of these developments and in view of recent significant progress in the control and manipulation of
quantum systems \cite{qiv} it is natural to extend these investigations into the quantum domain and
to explore characteristic phenomena of quantum networks.

%%%%%%%%%%%%%%%%%%%%%%%%%% brief characterization of quantum networks  %%%%%%%%%%%%%%%%%%%%%%%%%%%%%%%%%%%%%%%%%%%%
In a natural generalization of its classical analog
the nodes of a typical quantum network are formed by
spatially localized distinguishable quantum systems and
couplings between different nodes originating from
interactions or communication are described in general by a completely positive quantum operation \cite{quantop}.
Iterative applications of this quantum operation give rise to a
dynamical evolution of this quantum network.
So, in general contrary to typical interacting many-body quantum systems \cite{many-body} 
the couplings between different parts of a quantum network
cannot be described by infinitesimal generators, such as Hamiltonians or Lindblad operators.
Quantum networks can model rather different physical quantum systems, such as interacting gases or
in the context of quantum information processing a quantum-communication-based internet.
Contrary to their classical counterparts they are not only able to 
manipulate classical correlations but also to 
distribute entanglement which is a characteristic quantum resource. As
strong entanglement cannot be shared freely between
several quantum systems \cite{entanglement}
the distribution of entanglement
over a quantum network follows rules which are fundamentally different from those governing
the distribution of classical correlations.

%%%%%%%%%%%%%%%%%%%%%%%%%%%%%%%%%% MAIN AIM of this letter %%%%%%%%%%%%%%%%%%%%%%%%%%%%%%%%%%%%%%%%%%%%%%%%%%%%%%%%%%%%%
In this letter we explore characteristic features  of
the intricate interplay between entanglement and decoherence in quantum networks.
Whereas unitary evolution typically generates entanglement, i.e. genuine quantum correlations, decoherence tends to destroy them. Thus
the competition between these two counteracting tendencies is expected to produce interesting dynamical features.
%%%%%%%%%%%%%%%%%%%%%%%%%%%%%%%%%%%%%%%%%%%%%%%%%%%%%%%%%%%%%%%%%%%%%%%%%%%%%%%%%%%%%%%%%%%% 
In order to explore some of them let us consider the rather simple case of
quantum networks whose nodes are formed by elementary two-level quantum systems (qubits) coupled by randomly chosen
unitary controlled-NOT operations. In large quantum networks with large numbers of coupled nodes
this rather elementary random unitary dynamics gives already rise to complicated
dynamical features under repeated iterations. Even for moderately large numbers
of qubits (between $10$ and $30$) a detailed description 
becomes computationally intractable.
Therefore, alternative ways have to be developed. 

In the following it is shown that despite these complications the asymptotic dynamics of iterated random
unitary quantum operations 
exhibits very regular patterns. It
turns out that their asymptotic dynamics is governed by a typically low dimensional attractor space which is
determined completely by the unitary transformations involved
and which is independent of the probability
distributions with which these unitary transformations
are selected. This basic property of random unitary quantum operations allows us to investigate
systematically the asymptotic dynamics even of large quantum networks.
As an example we present first results exploring characteristic features of 
cyclic qubit networks whose dynamics is determined by random unitary controlled-NOT operations. 

In order to put the problem into perspective consider an arbitrary quantum system
whose dynamics is described by the iterated application of a random unitary operation 
\begin{eqnarray}
\Phi (\hat{\rho}) &=& \sum_{i\in I} p_i \hat{U}_i \hat{\rho}\hat{U}_i^{\dagger}.
\label{RUCH0}
\end{eqnarray}
Thus, a single application of this quantum operation
consists of selecting a unitary linear operator
$\hat{U}_i \in {\cal U} = \{\hat{U}_l\mid~l\in I\}$ 
randomly according to the normalized probability distribution $\{0<p_l\leq 1,~i\in I\}$
and of applying it onto
an initially prepared quantum state $\hat{\rho}\in {\cal B}({\cal H})$.
(${\cal B}({\cal H})$ denotes the Hilbert space of linear operators over the 
$d$-dimensional Hilbert space ${\cal H}$ describing
the quantum system under consideration.)
Correspondingly, after $n$ iterations of this random unitary operation the quantum system is in the state 
$\Phi^{(n)} (\hat{\rho})\equiv \Phi (\Phi (\Phi (\cdots (\Phi (\hat{\rho})))))$.

%%%%%%%%%%%%%%%%%%%%%%%%%%%%%%%%%%RUOP can describe various physical situations %%%%%%%%%%%%%%%%%%%%%%%%%%%%%%%
The dynamics of Eq.(\ref{RUCH0}) can model various physical situations. For example, if the quantum system consists
of a large number of distinguishable few-level systems it can model colliding
distinguishable low-dimensional systems whose collisions are well separated in time.
In this case each $\hat{U}_i \in {\cal U}$ 
describes the unitary evolution of a completed collision which occurs with probability $p_i$.
Alternatively the iterated dynamics of 
Eq.(\ref{RUCH0}) can also describe a quantum walk \cite{quantumwalk} whose ideal unitary dynamics is perturbed by random
imperfections. 
In the context of quantum information processing
this dynamics may model the probabilistic exchange of quantum information between 
different nodes within a quantum internet in which different nodes entangle each other by
unitary operations selected randomly by its users.

Random unitary operations
have  a number of characteristic properties. 
Obviously they belong to the class of unital quantum operations \cite{quantop,Bhatia2007} which
leave the maximally mixed quantum state
$\hat{\rho} = {\bf 1}/d^2$
invariant. Defining the
Hilbert-Schmidt scalar product $(\hat{A},\hat{B}) = {\rm Tr}(\hat{A}^{\dagger}B)$
in the Hilbert space of linear operators ${\cal B}({\cal H})$ 
the adjoint of $\Phi$ of Eq.(\ref{RUCH0}) is given by 
\begin{eqnarray}
\Phi^{\dagger} (\hat{\rho}) &=& \sum_{i\in I} p_i \hat{U}^{\dagger}_i \hat{\rho}\hat{U}_i.
\label{RUCH}
\end{eqnarray}
Thus,
in general
$\Phi$ is not normal, i.e. $[\Phi^{\dagger},\Phi] \neq 0$, so that it
cannot be
diagonalized. Nevertheless, its properties can still be analyzed systematically
with the help of Jordan normal forms \cite{linear_algebra}.
We show in the following that despite the resulting complications
the asymptotic dynamics of iterated random unitary operations $\Phi^{(n)}$
is governed by surprisingly regular patterns
which considerably simplify their description especially in the large-$n$ limit.
In special cases this asymptotic dynamics can even be determined analytically.

%%%%%%%%%%%%%%%%%%%%%%%%Determination of asymptotic dynamics %%%%%%%%%%%%%%%%%%%%%%%%%%%%%%%%%%%%%%%%%%%%%
For a discussion
of this asymptotic dynamics we use the fact that 
$\Phi$ of Eq.(\ref{RUCH0}) cannot decrease the von Neuman entropy $S$, i.e.
\begin{eqnarray}
S\left(\Phi (\hat{\rho})\right) &\geq& \sum_{i\in I} p_i S\left(\hat{U}_i \hat{\rho}\hat{U}^{\dagger}_i\right) = S(\hat{\rho}).
\label{RUCH11}
\end{eqnarray}
The resulting monotony of $S\left(\Phi^{(n)}(\hat{\rho})\right)$
and its boundedness for the finite dimensional quantum systems at hand imply that in the limit of large numbers of
iterations $\Phi^{(n)}(\hat{\rho})$
leads to a constant von Neuman entropy. Thus, 
the relations  \cite{OP}
\begin{eqnarray}
\hat{U}_i \hat{\rho}_{n}\hat{U}^{\dagger}_i = 
\hat{U}_{i'} \hat{\rho}_{n}\hat{U}^{\dagger}_{i'}
\label{asymptot1}
\end{eqnarray}
have to be fulfilled for all asymptotic quantum states
$\hat{\rho}_{n}= \Phi^{(n)}(\hat{\rho})$ with $n$ sufficiently large 
and for all unitary operations with $i,i'\in I$.
From Eq.(\ref{asymptot1}) we note that
the set of all its solutions 
forms a linear space, the attractor space ${\cal A} \subset {\cal B}$
and that $\Phi$ of Eq.(\ref{RUCH0})
acts on this attractor space unitarily.
As unitary transformations are normal
the restriction of
$\Phi$ onto the attractor
space ${\cal A}$ can be diagonalized
with the help of
a complete set of orthonormal eigenvectors
$\hat{X}_{\lambda} \in {\cal A}$. 
These
eigenvectors fulfill the eigenvalue equations and orthonormality constraints
\begin{eqnarray}
\hat{U}_i \hat{X}_{\lambda} \hat{U}^{\dagger}_i &=& \lambda \hat{X}_{\lambda},~
{\rm Tr}\left({X}^{\dagger}_{\lambda} {X}_{\lambda'}\right) = \delta_{\lambda \lambda'} 
\label{eigen}
\end{eqnarray}
for all $i\in I$. 
From Eq.(\ref{eigen})
it is apparent that the linear operators of the attractor
space ${\cal A}$ also form a $C^*$-algebra. This property implies a number of useful relations,
such as
\begin{eqnarray}
\mid \lambda \mid &=& 1,~
\hat{X}^{\dagger}_{\lambda} = 
\hat{X}_{\lambda^*},~
\lambda^n\lambda'^m \neq 1 ~\longrightarrow~{\rm Tr}\left(\hat{X}^n_{\lambda}\hat{X}^m_{\lambda'}\right) = 0.\nonumber
\end{eqnarray}
Thus, the attractor space ${\cal A}$ is defined by all linear operators $\hat{X}_{\lambda}$ fulfilling
Eqs.(\ref{eigen})
with the additional requirement $\mid \lambda \mid =1$. 
Furthermore, all asymptotically accessible quantum states $\hat{\rho}_n = \Phi^{(n)}(\hat{\rho})$
are contained in this attractor space, i.e.
$\lim_{n\to \infty}\hat{\rho}_n 
\in {\cal A}$.
Let us now introduce
the (continuous) projection operator
\begin{eqnarray}
{\cal P}~\left(.\right)~&=&
\sum_{\mid \lambda \mid =1}
{\rm Tr}\left(\hat{X}^{\dagger}_{\lambda}~\left(.\right)~\right)\hat{X}_{\lambda} 
\label{pop}
\end{eqnarray}
which projects the linear space ${\cal B}$
onto the attractor space ${\cal A}$. In view of the relation $1/\lambda^* = \lambda$ for $|\lambda | = 1$
it commutes with the random unitary operations of Eq.(\ref{RUCH0}), i.e.
$\left[{\cal P}, \Phi\right] = 0 $, so that
we can conclude with the help of Eqs.(\ref{eigen}) and (\ref{pop})
\begin{eqnarray}
&&
\lim_{n\to \infty}\Phi^{(n)}\left(\hat{\rho}\right)= 
{\cal P}\left(\lim_{n\to \infty}\Phi^{(n)}\left(\hat{\rho}\right)\right) 
=
\lim_{n\to \infty}\Phi^{(n)}\left(
\hat{\rho}_{out}
\right)=\nonumber\\
&&\lim_{n\to \infty}
\hat{U}_{l_0}^{(n)}
\hat{\rho}_{out}
\hat{U}_{l_0}^{(n)\dagger}
=
\lim_{n\to \infty}\sum_{\mid \lambda \mid =1}\lambda^n
{\rm Tr}\left(\hat{X}^{\dagger}_{\lambda}~\left(\hat{\rho}\right)~\right)\hat{X}_{\lambda}
\label{asymPhi}
\end{eqnarray}
with $l_0\in I$ arbitrary and with the projected initially prepared quantum state $\hat{\rho}_{out} = {\cal P}(\hat{\rho})$.
Eq.(\ref{asymPhi}) is a main result of this letter. It shows explicitly how the asymptotic dynamics of $\Phi$
of Eq.(\ref{RUCH0}) is determined by the structure of the attractor space ${\cal A}$.
Of course, 
Eq.(\ref{asymPhi}) could also have been derived in a direct but more cumbersome way with the help of a 
Jordan normal form decomposition of $\Phi$ \cite{NAJ}.  

According to Eq.(\ref{asymPhi})
the determination of the asymptotic dynamics of $\Phi^{(n)}$ 
can be divided into two main steps.
In the first step
one determines the set of eigenvalues $\lambda$ of unit modulus and the associated orthonormal basis $\{\hat{X}_{\lambda}\}$
of the linear space ${\cal A}$
with the help of Eqs.(\ref{eigen}).
Typically, this is a difficult task which may be facilitated by symmetries.
In the second step one evaluates the unitary asymptotic action of $\Phi^{(n)}$
onto the projected quantum state $\hat{\rho}_{out}$ according to Eq.(\ref{asymPhi}). 
Typically, the attractor space is expected to be of sufficiently low dimensions
so that Eq.(\ref{asymPhi}) implies significant simplifications. 

%%%%%%%%%%%%%%%%%%%%%%%%%%%%%%%% general consequences  %%%%%%%%%%%%%%%%%%%%%%%%%%%%%%%%%%%%%%%%%%%%%%%%%%%%%%%%%%%%%%%%%%%%%%%%%%%%%%%
Eq.(\ref{asymPhi}) 
hints at some remarkable general features of the asymptotic dynamics of $\Phi^{(n)}$. 
First of all,
the space of attractors ${\cal A}$
is determined completely by the set of unitary operators ${\cal U}$. In particular, 
this implies that the asymptotic dynamics is
independent of the classical probability distribution characterizing $\Phi$.
Nevertheless, in general this probability distribution 
still 
influences the rate of convergence. 
Furthermore,
if one of the unitary operations from the set ${\cal U}$ 
is  the unit-operation (apart from a global phase) the only possible
eigenvalue is given by $\lambda =1$.
The resulting asymptotic dynamics is thus stationary.
Another general consequence can be derived for unitary operations ${\cal U}$ which generate a ray-representation
of a finite group. Because the unit-element of this group is a product of the generators contained in ${\cal U}$,
for each eigenvalue $\lambda$
there 
is a natural number $n_{\lambda}$
with $\lambda^{n_{\lambda}}= 1$. Thus, the resulting
asymptotic dynamics is periodic. However, Eq.(\ref{asymPhi}) also applies to the most general non-stationary and non-periodic 
cases of asymptotic dynamics.

%%%%%%%%%%%%%%%%%%%%%%%%%%%%%%%%%%%%%% application to quantum network %%%%%%%%%%%%%%%%%%%%%%%%%%%%%%%%%%%%%%%
Let us now apply the general results of Eq.(\ref{asymPhi}) to the description of
the asymptotic dynamics of a particular class of quantum networks consisting of qubits which are
coupled by randomly selected controlled-NOT operations. 
A controlled-NOT operation 
between qubits $i$ and $j$ is defined by
$C_{i,j}|a\rangle_i\otimes |b\rangle_j =|a\rangle_i\otimes |b\oplus a\rangle_j.$
It is known to entangle qubits effectively because
it produces pure maximally entangled Bell states from separable states of the form
$(|0\rangle_{i} + |1\rangle_i)/\sqrt{2}\otimes|b\rangle_j$, for example.
Here, the pure states $|a\rangle$ with $a\in\{0,1\}$ denote orthonormal basis states of the computational basis
of a qubit and $a\oplus b$ with $b\in\{0,1\}$ denotes addition modulo $2$.
Let us further assume that 
our network has a one-dimensional cyclic topology so
that nodes $i$ and $i+1$ with $i=1,...,N$ are coupled by controlled-NOT operations $C_{i,i+1}$
and that in view of the cyclic topology
qubits $N+1$ and $1$ are identical. 

In order to determine the asymptotic limit of the corresponding random unitary operation one has to solve
the eigenvalue problem of Eqs.(\ref{eigen}). The possible eigenvalues can be determined easily by noting that
controlled-NOT operations have the property $C_{i,i+1}^2 = {\bf 1}$. Therefore, the only
possible eigenvalues of Eqs.(\ref{eigen}) are given by
$\lambda = \pm 1$. The determination of all
eigenvectors with eigenvalue $\lambda =1$ is facilitated by the observation that
the pure quantum states
$
|0\rangle$  and $|\Phi\rangle =\sum_{z=1}^{2^N-1}|z\rangle/\sqrt{2^N -1}$  
(with
$|z\rangle \equiv |j_N\rangle|j_{N-1}\rangle \cdots |j_2\rangle |j_1\rangle$,
$z = \sum_{i=1}^N 2^{i-1}j_i$,
and $j_i\in\{0,1\}$)
are invariant under all controlled-NOT operations of the cyclic network. Using the additional
fact that the unit operator is a solution of Eq.(\ref{eigen})
with eigenvalue $\lambda =1$ we find the following orthonormal eigenvectors for $\lambda =1$
\begin{eqnarray}
\label{attractor} 
\hat{X}_1 = |0\rangle\langle 0| &,& \hat{X}_2 = |0\rangle \langle \Phi|,~
\hat{X}_3 = |\Phi\rangle \langle 0| ,\\ 
\hat{X}_4 = |\Phi\rangle \langle \Phi|&,&
\hat{X}_5 = ({\bf 1} - |0\rangle \langle 0| - |\Phi\rangle \langle \Phi|)/\sqrt{2^N-2}.\nonumber
\end{eqnarray}
By a somewhat lengthy calculation 
one can prove from Eqs.(\ref{eigen}) by induction
that there are no additional eigenvectors.
Thus, for any number of qubits $N$
the eigenspace associated with $\lambda = 1$ is five dimensional and is given by Eq.(\ref{attractor}).
By induction one can also demonstrate that for $N > 2$ solutions of 
Eqs.(\ref{eigen}) with eigenvalue $\lambda = -1$ do not exist.
It is only in the special case of $N=2$ that a non trivial normalized eigenvector
exists. Explicitly it is given by \cite{NAJ}
\begin{eqnarray}
\hat{X}_6 &= -&|0\rangle|1\rangle\langle 1|\langle0| +|0\rangle|1\rangle\langle 1|\langle1| +|1\rangle|0\rangle\langle 0|\langle1|-\nonumber\\
&&|1\rangle|0\rangle\langle 1|\langle1| -|1\rangle|1\rangle\langle 0|\langle1| +|1\rangle|1\rangle\langle 1|\langle0|.
\label{two}
\end{eqnarray}

Having determined an orthonormal basis of the attractor space ${\cal A}$, the general form of the asymptotic dynamics
of the quantum network can be determined with the help of Eq.(\ref{asymPhi}).
Thus, for any number of qubits with $N > 2$ the projected quantum state $\hat{\rho}_{out}$ 
is given by
\begin{eqnarray}
\hat{\rho}_{out} &=& p \frac{\hat{P}_2 \hat{\rho} \hat{P}_2}{p} + (1 - p)
\frac{
{\bf 1} - \hat{P}_{2}
}{2^N-2}.
\label{asympstate}
\end{eqnarray}
This state
is stationary because it is invariant under
all controlled-NOT operations under consideration.
Here, the projection operator 
$\hat{P}_2 = |0\rangle \langle 0| + |\Phi\rangle \langle \Phi|$ projects onto the two-dimensional subspace
${\cal H}_2 \subset {\cal H}$ spanned by the pure states $|0\rangle$
and $|\Phi\rangle$ which are invariant under all controlled-NOT operations under consideration. The probability 
$p = {\rm Tr}\{\hat{P}_2 \hat{\rho}\}$ measures the overlap of the initially prepared quantum state $\hat{\rho}$ 
with this invariant subspace. 
Eq.(\ref{asympstate}) implies that any initially prepared quantum state $\hat{\rho}$ which is contained
completely in subspace ${\cal B}({\cal H}_2)$
is not affected by the randomly applied controlled-NOT operations. 
Thus, ${\cal B}({\cal H}_2)$ forms a decoherence-free subspace \cite{DFS,DFS1}. 

Let us finally discuss the convergence towards the asymptotic dynamics.
A convenient measure for the deviation between the quantum state after $n$ iterations $\hat{\rho}_n$
and the corresponding asymptotic quantum state $\hat{U}^{n}_{l_0}\hat{\rho}_{out}\hat{U}_{l_0}^{n\dagger}$
is the Hilbert-Schmidt norm or trace distance $D$
between both states with
$D^2 = 
\left(\hat{\rho}_n -\hat{U}_{l_0}^{n}\hat{\rho}_{out}\hat{U}_{l_0}^{n\dagger},
\hat{\rho}_n - \hat{U}_{l_0}^{n}\hat{\rho}_{out}\hat{U}^{n\dagger}_{l_0}\right)$.

\begin{figure}
\includegraphics[width=7.cm]{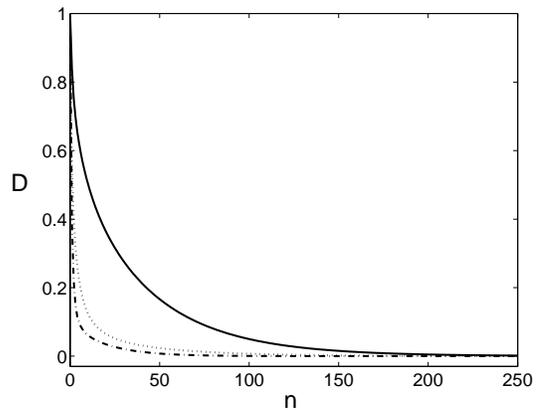}
\caption{Trace-norm distance $D$ between $n$-th iterate $\hat{\rho}_n$
and the projected quantum state $\hat{\rho}_{out}$
for different pure six-qubit states $\hat{\rho} = |\psi\rangle\langle\psi|$:
$|\psi\rangle = |000001\rangle~({\rm full}),~
|111111\rangle~({\rm dashdot}),~
|101010\rangle~({\rm dot}).$}
\label{Fig3}
\end{figure}

\begin{figure}
\includegraphics[width=7.cm]{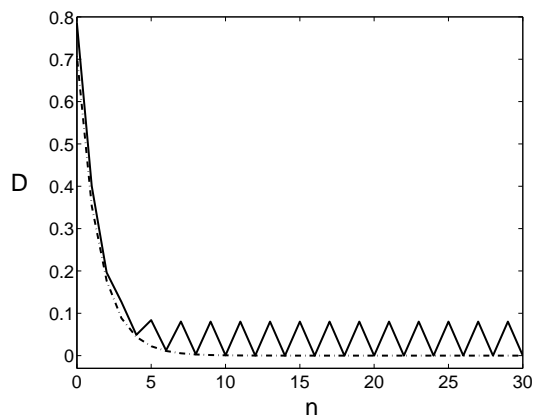}
\caption{Trace-norm distance $D$ 
between $n$-th iterate $\hat{\rho}_n$
and the projected quantum state $\hat{\rho}_{out}$ 
for different pure two-qubit states $\hat{\rho} = |\psi\rangle\langle\psi|/\sqrt{|\langle \psi|\psi\rangle|}$:
$|\psi\rangle = 0.3e^{i\pi/5}|10\rangle + e^{i\pi/7}|11\rangle~({\rm full}),~
|10\rangle + |11\rangle~({\rm dashdot}).$}
\label{Fig2}
\end{figure}

In Fig.\ref{Fig3} numerical results are presented depicting the
trace-distance $D$ between the states $\hat{\rho}_n$ and
$\hat{\rho}_{out} = \hat{U}_{l_0}^{n}\hat{\rho}_{out}\hat{U}^{n\dagger}_{l_0}$ 
for a cyclic quantum network consisting of $N=6$ qubits
and for various initial quantum states $\hat{\rho}$ with a uniform probability distribution.
It is apparent that convergence is achieved in a strictly monotonic way. 
Nevertheless, the rate of convergence depends on the initially prepared quantum state.

In Fig.\ref{Fig2} analogous numerical results are depicted for a cyclic quantum network consisting of
two qubits and for the 
trace-distance $D$
between the states $\hat{\rho}_n$ and $\hat{\rho}_{out}$.
Contrary to the previously considered case,
now the asymptotic dynamics is not always stationary, i.e. typically $\hat{\rho}_{out} \neq
\hat{U}_{l_0}^{n}\hat{\rho}_{out}\hat{U}^{n\dagger}_{l_0}$, because according to Eq.(\ref{two}) there is also
a non trivial eigenspace with eigenvalue 
$\lambda = -1$. Depending on whether the initially prepared quantum state $\hat{\rho}$ overlaps with this eigenspace
or not the asymptotic dynamics of $\hat{\rho}_n$ is non-stationary or stationary.

In conclusion, we have presented a general method which allows us to determine the asymptotic dynamics of
quantum networks under the influence of iterated random unitary operations.
It is based on the determination of the associated 
asymptotic attractor space.
This attractor space 
is independent of the probability distribution involved and is
typically low dimensional thus simplifying the asymptotic dynamical description considerably.
The results presented for cyclic qubit networks
demonstrate explicitly how the properties of the attractor space determine
the resulting global entanglement of the quantum network. It is expected that
the exploration of
the structure of attractor spaces 
associated with the dynamics of more general networks
will 
shed light also onto other open problems of network dynamics in the quantum domain, such as the connection between 
network topology and the resulting asymptotic dynamics. 

Financial support by the DAAD, by the Alexander von Humboldt Foundation, and by MSM6840770039 and MSMT LC06002 of the Czech Republic
is acknowledged.


\begin{thebibliography}{99}
\bibitem{Barabasi} R. Albert and A.L. Barabasi, Rev. Mod. Phys. {\bf 74}, 47 (2002).
\bibitem{qiv} D. Bru\ss\ and G. Leuchs (eds.), \emph{Lectures on Quantum Information} (Wiley-VCH, Weinheim, 2007).
\bibitem{quantop}
K. Kraus, \emph{States, Effects and Operations, Fundamental Notions of Quantum
Theory} (Academic, Berlin, 1983). 
\bibitem{many-body}
B. Sutherland, 
\emph{Beautiful Models: 70 Years of Exactly Solved Quantum Many-Body Problems}
(World Scientific, Singapore, 2004). 
\bibitem{entanglement} 
V. Coffman, J. Kundu, W. K. Wootters, Phys. Rev. A {\bf 61}, 052306 (2000).
\bibitem{quantumwalk} 
N. Konno, \emph{Quantum Walks}, Lecture Notes in Mathematics {\bf 1954}, 309 (2008).
\bibitem{Bhatia2007}
R. Bhatia, \emph{Positive Definite Matrices} (Princeton UP, Princeton, 2007).
\bibitem{linear_algebra}
D. T. Finkbeiner II, \emph{Introduction to Matrices and Linear
Transformations}, third Edition (Freeman, San Francisco, 1978).
\bibitem{OP} M. Ohya and D. Petz, \emph{Quantum Entropy and Its Use} (Springer, Berlin, 2004).
\bibitem{NAJ} J. Novotn\'y, G. Alber, I. Jex (in preparation).
\bibitem{DFS} 
P. Zanardi and M. Rasetti, Phys. Rev. Lett. {\bf 79}, 3306 (1997).
\bibitem{DFS1} D.W. Kribs, Proceedings of the Edinburgh Mathematical Society {\bf 46}, 421 (2003).
\end{thebibliography}
\end{document}